\documentclass[12pt,preprint]{aastex}
\def\av#1{\langle #1\rangle}

\newcommand{\swifts}{Swift~J1753.5$-$0127 {}}
\begin{document}
\title{X-ray Spectral and timing properties of the black hole x-ray transient
 Swift~J1753.5~--~0127}
\author{G. B. Zhang\altaffilmark{1,2},  J. L. Qu\altaffilmark{1}, S. Zhang\altaffilmark{1},
 C. M. Zhang\altaffilmark{3}, F.
Zhang\altaffilmark{1}, W. Chen\altaffilmark{1}, L. M.
Song\altaffilmark{1}, S. P. Yang\altaffilmark{2} }
\affil{zhanggb@mail.ihep.ac.cn, qujl@ihep.ac.cn,
szhang@mail.ihep.ac.cn,
 zhangcm@bao.ac.cn, zhangfan@mail.ihep.ac.cn,
chenw@mail.ihep.ac.cn, yangship@hebtu.edu.cn} \affil{$^1$Laboratory
for Particle Astrophysics, Institute of High Energy Physics, Beijing
100049,  China} \affil{$^2$Hebei Normal University, Shijiazhuang
050016,  China} \affil{$^3$National Astronomical Observatories,
Chinese Academy of Sciences, Beijing 100012,  China}

\begin{abstract}
We have carried out detailed analysis on the black hole candidate
(BHC) X-ray transient {\swifts}  observed by the Rossi X-Ray Timing
Explorer ({\sl RXTE}) during its outburst in 2005~--~2006. The
spectral analysis shows that the emissions are dominated by the hard
X-rays, thus revealing the low/hard state of the source during the
outburst. The peak luminosity is found lower than the typical value
of balancing  the mass flow and  evaporation of the inner edge of
disk \citep{meyer04}. As a result, the disk is prevented  from
extending inward to produce strong soft X-rays,  corresponding to
the so-called high/soft state. These are the typical characteristics
for a small subset of BHCs, i.e. those soft X-ray transients stay at
the low/hard state during the outburst. In  most observational time,
the QPO frequencies are found to vary roughly linearly with the
fluxes and the spectral indices, while the  deviation from this
relationship at the peak luminosity might provide the first
observational evidence of a partially evaporated inner edge of the
accretion disk. The anti-correlation between the QPO frequency and
  spectral color suggests that the global disk oscillation model
  proposed by \cite{titarchuk00} is not likely at work.
\end{abstract}


\section{INTRODUCTION}

Galactic black hole binaries are classified by their X-ray features,
such as the strength and temperature of the soft multi color
blackbody component, the hard X-ray emission, the X-ray luminosity
and   timing properties \citep{mitsuda84,vander95,remillard06}.
Black hole X-ray transient outbursts are known to pass through a
number of X-ray spectral states over the course of their outbursts;
these spectral states are also associated with the specific jet
behaviors \citep{vander95,vander00,vander05,mcclintock06}. Several
states (very high state, high/soft state, intermediate state,
low/hard state, quiescent state etc.) have been identified to
characterize black hole accretion
\citep{tanaka95,vander94,poutanen99,mcclintock06}. The soft state,
commonly described as $\sim 1$ keV thermal emission, was usually
observed when the source was   at higher luminosity, thereby
prompting the name``high/soft state''. The hard state, with a
typical photon index $\Gamma \sim 1.7$, is generally observed when
the source was in low luminosity, hence named "low/hard state". The
quiescent state, appearing at low luminosity levels, characterizes
the long periods of quiescence of transient systems. In the very
high state, black hole becomes exceedingly bright ($L_x$ $\geqslant$
0.2 $L_{Edd}$), and its X-ray spectrum displays a substantial
nonthermal radiation, which may constitute 40--90\% of the total
flux. In such a case, the photon index is typically $\Gamma \geq
2.4$, which is steeper than  that ($\Gamma \sim 1.7$) observed  in
the hard state \citep{mcclintock06}. More than 130 soft X-ray
transients (SXTs) have been found \citep{brocksopp04} up to now, and
their number is growing every year. In particular, there are a
number of supposed black hole SXTs which do not show a soft
component in their X-ray spectra. Instead they remain  in the
low/hard state (the X-ray emission was dominated by the power-law
spectra) throughout the outburst \citep{brocksopp04}. In addition,
there are very similar X-ray spectral properties in this subset, but
the light curve morphologies of these sources are very different,
including the X-ray, optical and radio light curves. As declaimed,
{\swifts}, a newly detected black hole X-ray transient, likely
belongs to the subset of BHCs \citep{miller06}.

A new, bright, variable  $\gamma$-ray and X-ray transient source,
{\swifts}, was detected firstly by the {\it Swift} BAT on  May 30,
2005 \citep{palmer05}, then its position,  $R.A.=17^h53^m28.3^s$,
$Dec.=-1^{o}27^{'}9.3^{''}$ (J2000.0) with an estimated 90\%
confidence uncertainty of $6^{''}$, was given by X-ray
Telescope(XRT) \citep{burrows05}. {\swifts} was also clearly
detected in soft X-ray with the {\it Swift} X-ray Telescope (XRT)
 \citep{morris05}. Later on, its optical counterpart   was observed
by the MDM 2.4 m telescope on July 2, 2005 \citep{halpern05}, and
its radio counterpart was detected at 1.7 GHz with the Multi-Element
Radio-Linked Interferometer Network (MERLIN) on  July 3, 2005.
Further radio observations on July 4 and 5 indicate that this source
is variable, strengthening the likely association with the X-ray
transient \citep{fender05}. These radio observations suggest that
the source might in its low/hard state during the outburst.

To fully understand the BHC behavior  of X-ray transient, it is
instructive to outline the source properties for the whole  range of
X-ray flux from the maximum of the outburst to quiescence, so as for
this purpose {\sl RXTE} has proved to be an extremely unique tool
for studying the transients. Using the  {\sl RXTE} and {\sl
XMM-Newton} data, \cite{miller06} have analyzed the spectra of
{\swifts} when the source approached the quiescence state. Their
results tend to support a cool (kT $\sim 0.2$ keV) accretion disk
around the central compact object, which may locate  at/near  the
innermost stable circular orbit (ISCO). Furthermore, the detection
of   a 0.6 Hz quasi-periodic oscillation (QPO) by {\sl RXTE} and the
measurement of a hard power-law spectrum by {\sl Swift/XRT}
\citep{morris05} suggest a hard state for this source
\citep{morgan05} during the outburst.

In this paper, we have analyzed the  TOO observations carried out by
Proportional Counter Array (PCA) and High-Energy X-ray Timing
Experiment (HEXTE) onboard {\sl RXTE}, and investigated in detail
the evolutions of the outburst, with the focus on the spectral and
the timing properties. The paper is organized to show the
observations and data reduction in Sec.2, the presentation of the
main results in Sec.3, and finally the discussion and conclusion are
performed in Sec.4 and Sec.5.

\section{OBSERVATION AND DATA REDUCTION}
Onboard the {\sl RXTE} satellite there are three detectors. The PCA
and HEXTE are the two co-aligned spectrometers with  narrow fields
of view. They provide a broad energy coverage from 2 to $\sim$ 250
keV. Another detector, the so-called ASM, is used to track the
long-term behavior of the source in the energy band 2-13 keV. The
PCA consists of 5 non-imaging Xe multiwire proportional counter
units (PCUs). Each unit covers the energy  2-60 keV and has the time
resolution $\sim 1\,\mu{\rm s}$ ($2^{-20}$ s). PCA has
$1^o\times1^o$ field of view and provides in sum the collecting area
of $\sim 6500\,{\rm cm}^2$. The HEXTE comprises two clusters, each
with 4 scintillation detectors sensitive to photons in the range
15-250 keV, collimated to view a common $1^o$ field, and these eight
detectors in the clusters provide a total collecting area of
1600$cm^2$ \citep{gruber96}.

The {\sl RXTE} TOO data of {\swifts} are available for observations
from July 6, 2005 to March 11, 2006 (Fig.1). The software package
HEASOFT version 6.0 was adopted for reduction of the PCA data. We
get rid of the data for which the angle of the source above the
Earth limb is less than 10 degrees or for which the pointing offset
is greater than 0.02 degrees. We take the Standard-2  mode data ( 16
s time resolution and 129 energy channels between 2-60 keV) for
energy spectral analysis,  the Event mode E\_125us\_64M\_0\_1s
($\sim122$ $\mu$s time resolution and 64 energy channels between
2-60 keV) and the Standard-1 mode (0.125 s time resolution and 1
energy channel between 2-60 keV) for timing analysis. Only
proportional counter units (PCUs) 0 and 2 were always active during
our observation, but the propane layer on PCU-0 was lost when a
micro-meteorite created a small hole in the front window of PCU-0.
Therefore, we take the Standard-2 events from the PCU-2 and extract
the energy spectra from the upper xenon layer.

The extracted PCA spectrum covers the 2-60 keV
bandpass in 129 channels, each sampled in every 16 seconds. The
energy spectra at 18-180 keV  are derived from the HEXTE-A data.
Because the detector 2 of the HEXTE cluster B   lost the spectral
capability, only the  data from cluster A are
analyzed. We use the software package XSPEC, version 12.2.1, to
analyze the broadband (3-180 keV) spectra. The background spectra
are made by the FTOOLS {\em pcabackest}, with the latest `bright
source' background model enclosed. An instrument response files are
obtained with the tool {\em pcarsp}. An additional  0.5\% systematic
errors are added to the PCU-2 spectra because of the calibration
uncertainties.

\section{ANALYSIS and RESULTS}

\subsection{ASM and PCA Light Curves, Color-Color Diagram}
\begin{figure}[pthpthpth]
\includegraphics[width=6in,height=6.5in,angle=-90]{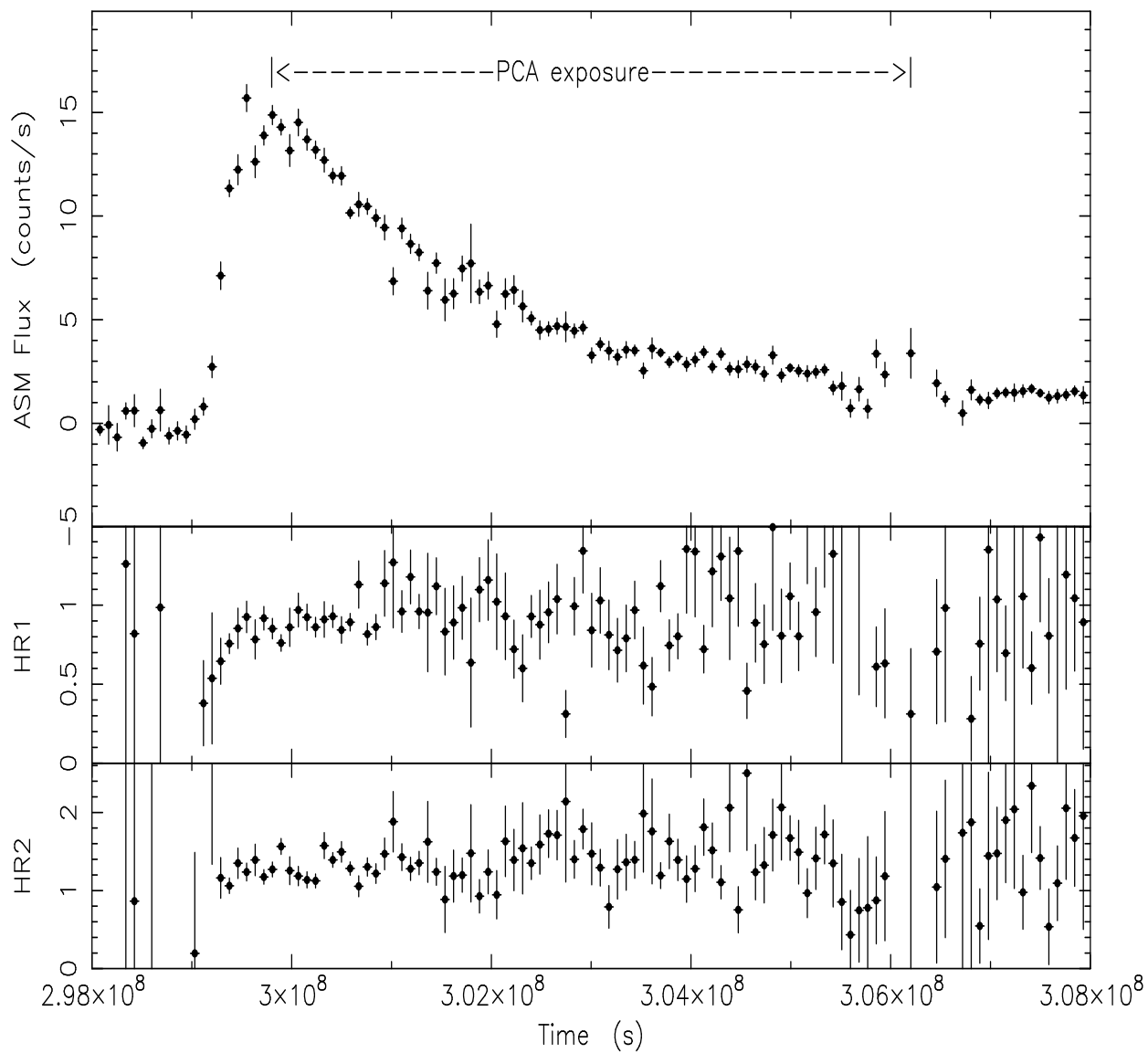}
\caption{The  ASM light curve in the 1.3--12.1 keV energy band of
{\swifts} during the 2005 outburst. Each bin is averaged over one
day. On top of this light curve shows the time period where PCA
observations are available in the archive. The lower two panels give
the soft color HR1 (3.0-5.0 keV/1.3-3.0 keV) and the hard color HR2
(5.0-12.1 keV/3.0-5.0 keV).}\label{asmlc}
\end{figure}

 ASM monitored the light curve of the entire outburst in the
1.3--12.1 keV energy band. As shown in  Fig.\ref{asmlc} is the ASM
light curve with each bin averaged over one day. The outburst
started on June 29, 2005, reached the maximum  of $\sim 210$ mCrab
(1 Crab $\sim$ 75 counts/s
\footnote{http://xte.mit.edu/ASM\_lc.html}) on July 06, 2006, and
finally decayed  exponentially to quiescence. The fast-rise and
exponential-decay (hereafter FRED) light curve is considered as the
`classic' signature of a soft X-ray transient \citep{chen97}. To
investigate the  evolution of the energy spectrum, two colors are
investigated (see, e.g. \cite{hansiger89}): the soft color is the
count ratio of 3.0--5.0 keV to  1.3--3.0 keV, and the hard color is
the count ratio of 5.0--12.1 keV to 3.0--5.0 keV. In Fig.\ref{asmlc}
the spectral evolution is  obvious only for soft color during the
formation of the outburst.

As shown in panel A of Fig.\ref{lc_color_qpo_time} is the available
PCA observations of {\swifts} during the decay of outburst. The
residuals resulted from  fitting this light curve with two
exponential functions  indicate either the fluctuation of the flux
or several  additional small bursts. The light curve is subdivided
into three parts according to the evolution of the  colors and the
OPQ, as will be shown in the following.

The PCA colors are defined as the soft ( 6.1-9.4 keV/2.1-5.7 keV)
and the hard ( 13.5-16.9 keV/9.8-13.1 keV). Both colors increase
linearly when the burst  decay from the flux maximum to where the
linear relationship becomes rather weak, and the corresponding  time
period  is defined  as the  time zone 'part 1' (panel C/D of
Fig.\ref{lc_color_qpo_time}). Later on, both colors keep likely
constant during the rest of the decay. This trend is rather
straightforward  in the color-color diagram ( Fig.\ref{coco}).
Similar linear relationship between the  soft  and hard color was
reported in GRS 1915 + 105  by \citep{belloni00} in its low/hard
state. The second time zone (part 2) is ended when the QPO signal
vanishes. One sees from the Fig.\ref{lc_color_qpo_time} that in the
third time zone (part 3) the QPO signal appears occasionally around
the time when  the source is likely to have additional flare in flux
and evolution in hard color.

\subsection{Timing Analysis}

The power density spectra are produced with FTOOLS {\em powspec} for
the Event mode (E\_125us\_64M\_0\_1s) data in the energy band 2-60
keV. As shown in Fig.\ref{qpo_PDS} is one sample of the  PDS
obtained from the PCA observation ID 9143-01-01-00. The PDS consists
of several components  typical for the low-hard state in accreting
BHs \citep{cui99}: the   white noise at the low frequencies and the
red noise at the high frequencies. These noises are generally fitted
by a broken power law function  and the QPO in PDS is fit with a
Lorentzian function. The errors are estimated by  setting $\Delta
\chi^2$ = 1. For {\swifts}, 39 of 62 PCA observations are found to
have low-frequency quasi-periodic oscillations (LFQPOs) at the
frequencies $\sim$0.1--0.9 Hz (see Table \ref{qpotable}). These
results are obtained without taking into account  the harmonics in
the spectral fitting. The LFQPOs are a common feature for the X-ray
BH binary while staying in their low/hard state, and elsewhere
detected already in V404 Cyg, GRO J1719-24, GRO J0422+32, GS 1354-64
and XTE J1118+480 \citep{brocksopp04}.

During the decay of the outburst, the QPO has a rough trend of
evolving to the lower frequencies (Fig.\ref{lc_color_qpo_time}).
Such a trend is obvious in the plot of the flux against  QPO
frequency (Fig.\ref{qpoflux}), where   a  linear relationship is
well established. Similar feature was reported in the other BHC like
GRS 1915 + 105 \citep{markwardt99,reig00}. It is worth noting that
 several QPO frequencies in Fig.\ref{qpoflux} strongly deviate
from the overall linear relationship. These QPOs are detected  when
the outburst either reached to the flux maximum or decayed to rather
low flux levels.

\subsection{Spectral Analysis}

The energy spectra of the black hole X-ray binaries are generally
described by the joint components of a blackbody and a power-law
shape\citep{remillard06}. In such a model the soft X-rays are
thought to be produced in the accretion disk while the hard X-rays
from the inverse Compton scattering of the soft photons off the
electrons in the hot corona residing in the inner disk.

To find the proper components in the energy spectrum we take the
{\sl RXTE} observation 91432-01-01-00 for spectral fitting. This
observation was carried out on July 7, 2005 (indicated in
Fig.\ref{lc_color_qpo_time} by the arrow symbol), and the data have
almost the best statistics. After standard screening, the remained
exposures for PCA and HEXTE are  6.4 ksec and 2.1 ksec,
respectively. The {\sl RXTE} PCU-2 (3.0--23.0 keV) and HEXTE-A
(18.0--180.0 keV) spectra are combined. A scaling factor is
introduced to account for the different normalization between PCA
and HEXTE. This parameter is set to unit for PCA and free for HEXTE.
The hydrogen column density is fixed at  $\rm N_H$ =$1.6\times
10^{21} cm^{-2}$. To account for the major component in the energy
spectrum, we have performed several trials. The fitting   results in
a reduced $\chi^2$ $\sim$ 10.9 with a simple power law model, a
reduced $\chi^2$ $\sim$ 6.14 with a compTT model and a reduced
$\chi^2$ $\sim$ 5.10 with a broken power-law model (see
Fig.\ref{phamodel}). Therefore we take for the first step the broken
power-law model. For the improvement of fittings, the excesses
showing up in the residuals of Fig.\ref{phamodel}  require the extra
models. By adding  a disk blackbody model ("diskbb"in XSPEC) the fit
is improved with the reduced $\chi^2$ $\sim$ 2.21. The excess
residing at the energies between 5 and 8 keV could in principle be
the contribution from the disk reflection.  By taking the smeared
edge model (\cite{ebisawa94}, ``smedge''in XSPEC) into account, the
reduced $\chi^2$ goes down to 0.97. However, the parameters of disk
blackbody and smeared edge cannot be constrained. A substitution of
the smeared edge model by a Gaussian with the center fixed at 6.4
keV results  in  the reduced $\chi^2$ of 1.16, but the errors for
disk blackbody parameters are not constrained as well by the data.
Accordingly,  the disk blackbody model is replaced by a blackbody
model (``bbodyrad'' in XSPEC). This gives a reduced $\chi^2$ of
1.19, and the parameters are well constrained as the follows: the
blackbody temperature kT = $0.66\pm 0.04$ keV, normalization $\rm
Norm_{bb}$ = $330^{+140}_{-100}$, equivalent blackbody radius
$21.3^{+4.2}_{-3.4}$ (d/8.5  kpc) km, break energy BreakE = $36.4^{+
1.8}_{- 1.7}$ keV, photon index $\Gamma_1$ = $1.64\pm 0.01$ and
$\Gamma_2$ = $2.04\pm 0.03$, equivalent width of Fe K$\alpha$ line
EW = 0.152 keV,  an unabsorbed flux of $4.7\times 10^{-9} \ \rm erg
\ cm^{2}\ s^{-1}$(2-10 keV) and a luminosity of $4.1\times 10^{37}\
\rm erg \ s^{-1}$ $(d/\rm8.5\ kpc)^2$ (or $L_X/L_{Edd} = 3.1\times
10^{-2} \ (d/8.5\ \rm kpc)^2 \ (M/10M_\odot )$). The comparison of
the trials with different models
 refers to Table \ref{tab:model}. By applying the
bbodyrad+bknpower+gauss model (hereafter BBG model), the whole data
are analyzed and the results are presented in Table \ref{tab:bknpo}.

\section{DISCUSSION}
Comparing the light curve morphology of the 2005-2006 outburst of
{\swifts} with that of the 2000 first outburst of XTE J1118+480
\citep{wood01}, we find that  both exhibit  the  very similar
FRED-shaped form, which is often considered as a characteristic for
the `soft X-ray transients'(SXTs) \citep{lasota01}, then the account
for the  SXT outbursts is usually  referred to  the thermal-viscous
disk instability model (DIM) \citep{lasota01,brocksopp04} or to a
diffusion model \citep{wood01}. The FRED-shaped light curve of
{\swifts} can be well described using above two models.

The high and low spectral states firstly identified in Cyg X-1
\citep{tananbaum} have   been observed in a number of X-ray
binaries, where the ``high'' and ``low'' terminology is originally
chosen based on the 2-10 keV X-ray flux. It is later found that the
low state corresponds to a hard spectrum and the high state to a
soft spectrum. The spectra and fluxes (see Table \ref{tab:bknpo})
indicate that {\swifts} is in its low/hard state. Strikingly, the
observed state of {\swifts} is similar to those of a few BHCs (for
instance, XTE J1118+480 and GRO J0422+32), remaining in the low/hard
state during outburst.  For explaining the behavior of the SXTs
which can remain in the low/hard state during outburst, a model was
proposed by \cite{meyer04}. In their model, a transition to the
high/soft state depends on the peak mass flow rate in outburst, and
the critical mass flow rate range is 0.02 to 0.05 $L_X/L_{\rm Edd}$.
The thermal component comes from the blackbody emission of a
truncated thin disk and the non-thermal one from the Comptonization
of the soft X-ray. The highest flux of \swifts is $L_X/L_{Edd} =
3.1\times 10^{-2} \ (d/8.5\ \rm kpc)^2 \ (M/10M_\odot )$, which lies
within the declaimed critical mass flow rate. We also find that the
thermal component  presented in the spectra from the outburst to
quiescence  might come from the disk. But the parameters of the
thermal component cannot be constrained by the PCA  data.
\cite{miller06} analyzed the {\sl XMM-Newton} and {\sl RXTE}
observations when the source is near its quiescence. A thermal
component which comes from a thin disk was also found, however, a
smaller disk radius is not consistent with Meyer-Hofmeister's model.
\cite{miller06} suggest that the {\swifts} may have a low mass black
hole than the typical black hole candidates. The low mass black hole
has a less evaporation efficiency to allow the disk to grow near the
ISCO, which  accounts  for the thermal component\citep{meyer04}.

\begin{figure}[ptbptbptb]
\includegraphics[width=6.3in,height=6.3in,angle=0]{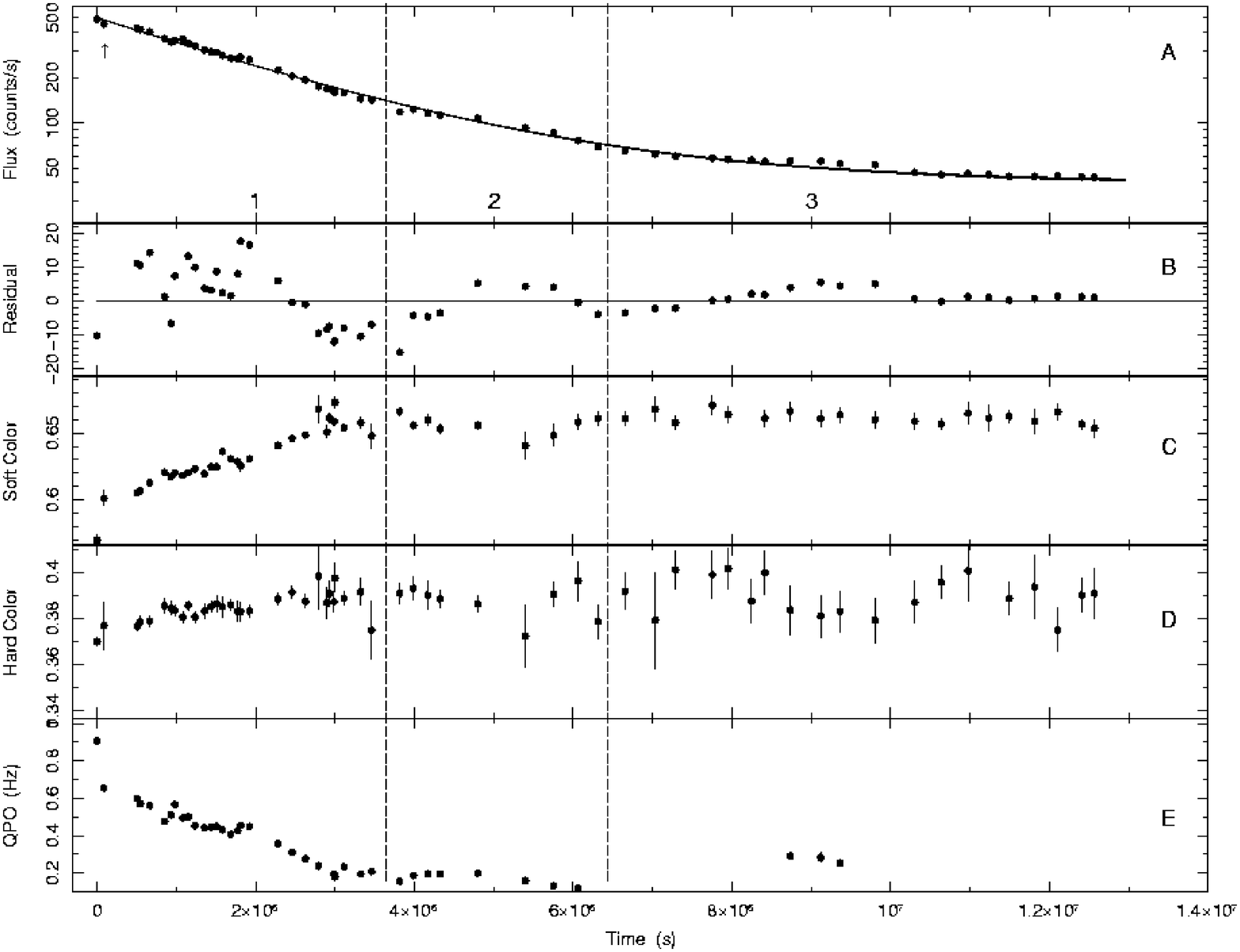}
\caption{ The time evolutions of the flux (A), spectral color (C):
soft color of 6.1-9.4 keV/2.1-5.7 keV; (D): hard color of 13.5-16.9
keV/9.8-13.1 keV) and QPO frequency (E). The solid line in panel A
shows the fit of the light curve and the residuals are shown in
panel B. }\label{lc_color_qpo_time}
\end{figure}

\begin{figure}[ptbptbptb]
\includegraphics[width=5in,height=6in,angle=-90]{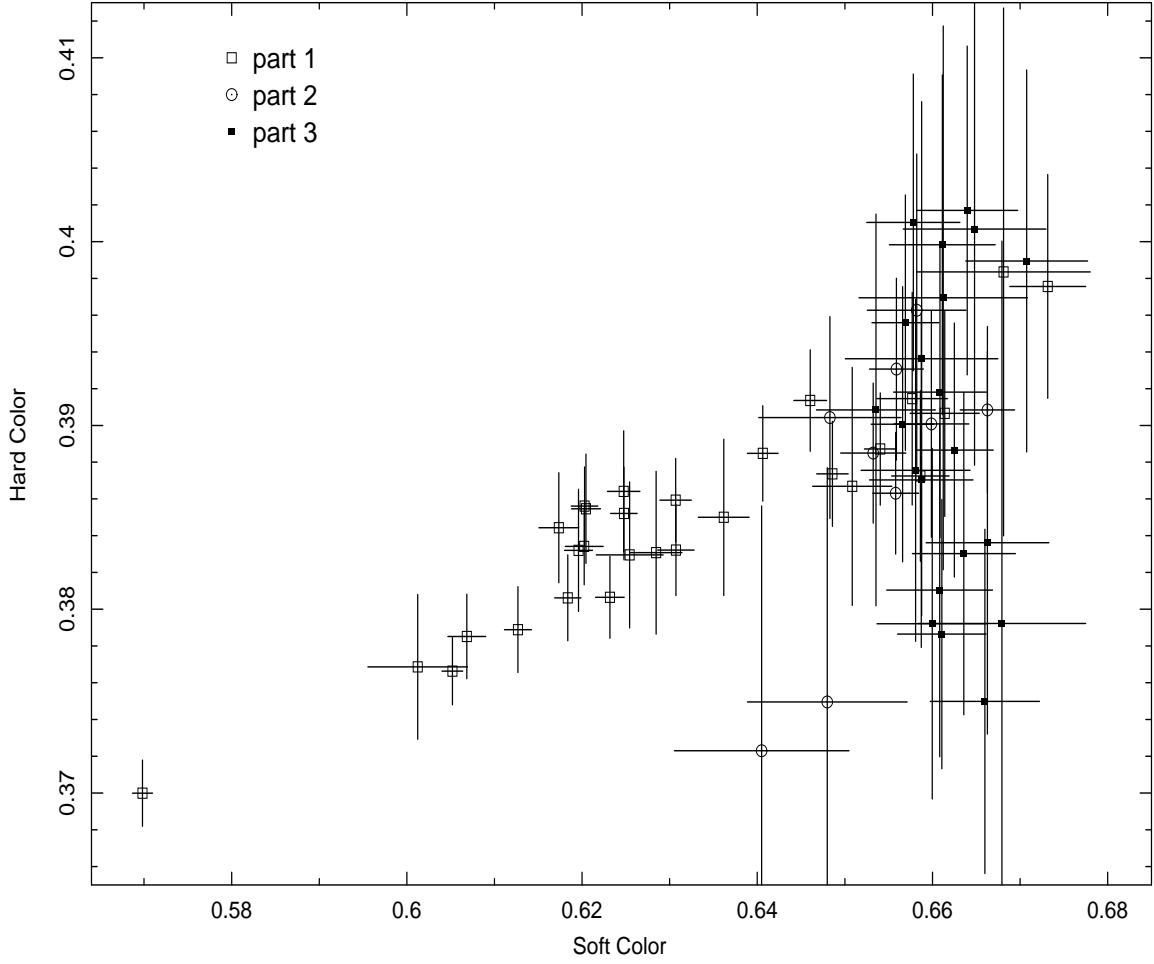}
\caption{ Background corrected color-color diagram of {\swifts}. The
soft color and the hard color have the same definitions as in Fig.2.
}\label{coco}
\end{figure}

\begin{figure}[ptbptbptb]
\includegraphics[width=5in,height=6in,angle=-90]{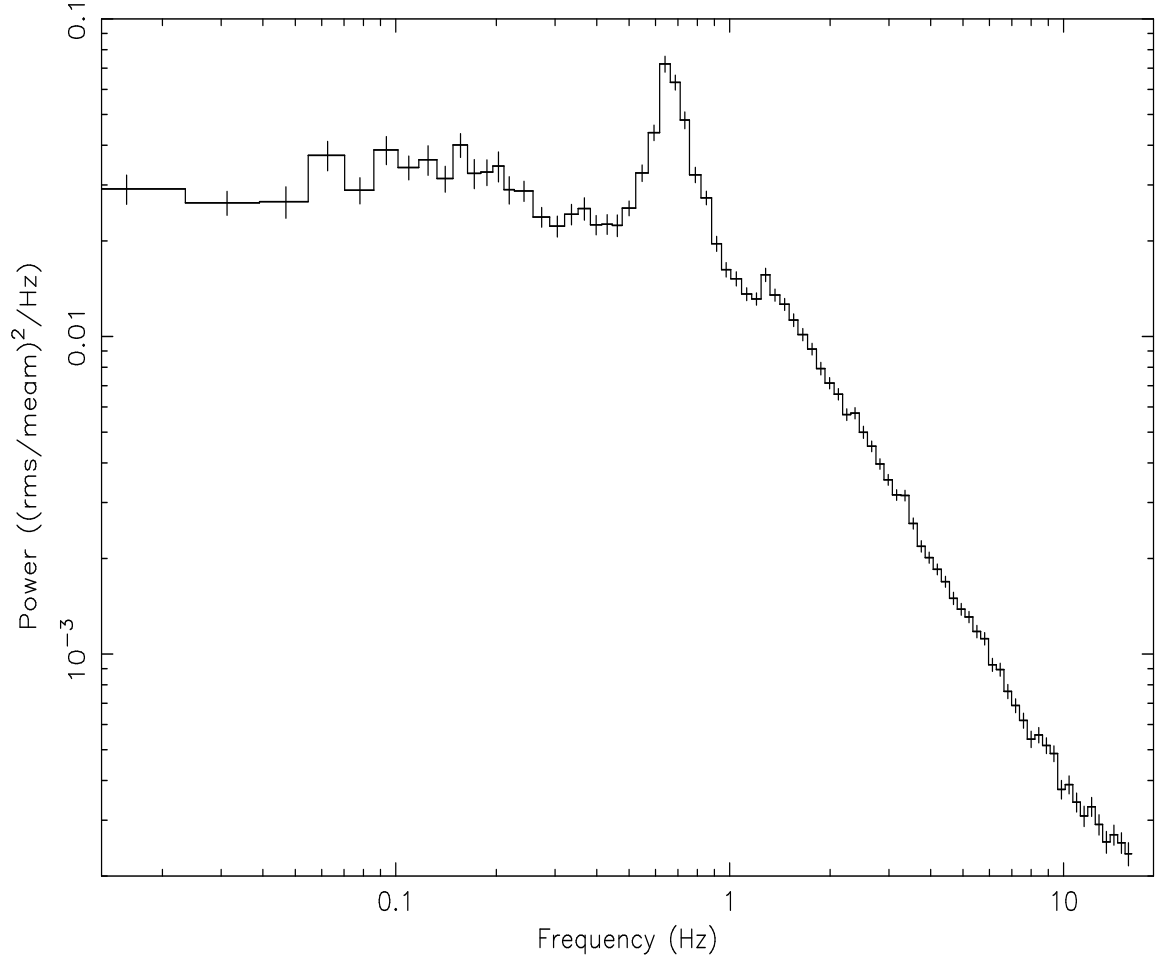}
\caption{ A typical PDS of {\swifts}, where a 0.6 Hz QPO shows up.
The corresponding observation is ID 9143-01-01-00.
 } \label{qpo_PDS}
\end{figure}

The X-ray PDS of {\swifts} displays LFQPOs in the range 0.1 to 0.9
Hz (see panel E of Fig.\ref{lc_color_qpo_time}), which is
extraordinarily important for understanding  the behaviors of
accreting black holes \citep{vander05,belloni02}, and usually they
are associated with the nonthermal or low/hard states and
transitions, implying the QPO emitting regimes to be far away from
ISCO, perhaps $\sim$100 Schwarzschild radius for the $\sim$10 solar
mass central object if the Keplerian frequency (maybe not necessary)
is taken into account. Two models are often mentioned to describe
the LFQPO \citep{brocksopp04}. \cite{titarchuk00} suggest that the
LFQPO of X-ray transients may be caused by the vertical oscillation
of the inner edge of accretion disk, and they predict a proportional
correlation between QPO frequency and the color parameter.
Furthermore, they also suggest  that the spectrum gets harder and
the QPO frequency increases when the disk mass accretion rate drops.
However, our results would expect the inner edge of disk to move
outwards with decreasing the QPO frequency, while moving back to its
``quiescent state'', which has been hinted in panel E of
Fig.\ref{lc_color_qpo_time}. With the exponential-decay light curve,
the frequencies of LFQPOs are decreasing, as shown in
Fig.\ref{lc_color_qpo_time}. So, generally, there exists an
approximate linear relation between the QPO frequency and  flux, and
the similar phenomena have been noted in GRS 1915 + 105
\citep{markwardt99,reig00}. In Fig.\ref{qpo_color}, a plot of the
soft/hard color versus the QPO frequency, we can find that both the
soft and hard colors drop with the QPO frequency, and a similar
anti-correlation between QPO frequency and hardness ratios has been
found in  BHC GRS 1915+105 \citep{reig00}. Therefore, LFQPOs of
{\swifts} might be unsatisfactorily explained by the model of
\cite{titarchuk00}.

\begin{figure}[ptbptbptb]
\includegraphics[width=5in,height=6in,angle=-90]{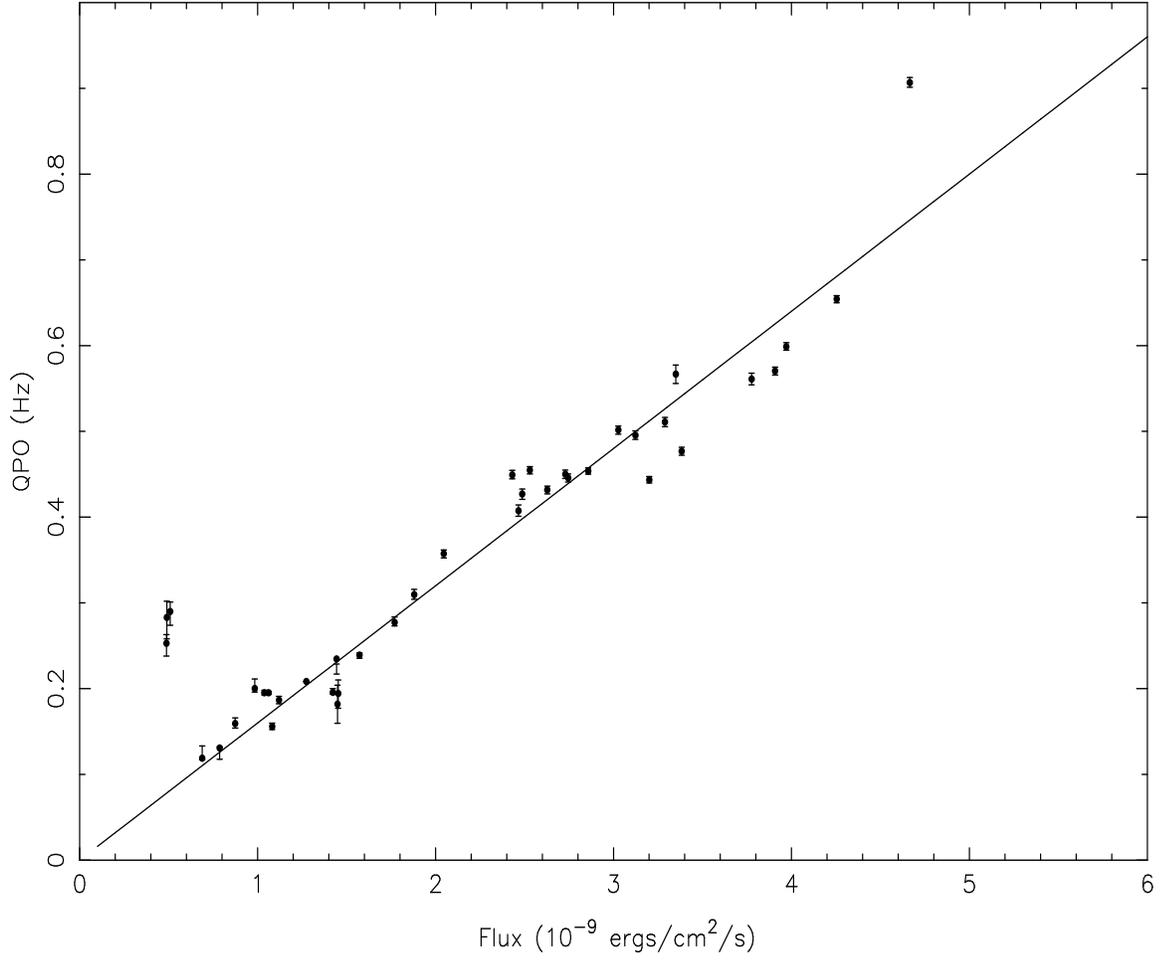}
\caption{ The correlation between flux and QPO frequency and the
linear fit. Three data points at the lowest fluxes and one at the
peak flux are not enclosed in the fitting procedure.
 } \label{qpoflux}
\end{figure}

\begin{figure}[ptbptbptb]
\includegraphics[width=6.3in,height=6.11in,angle=0]{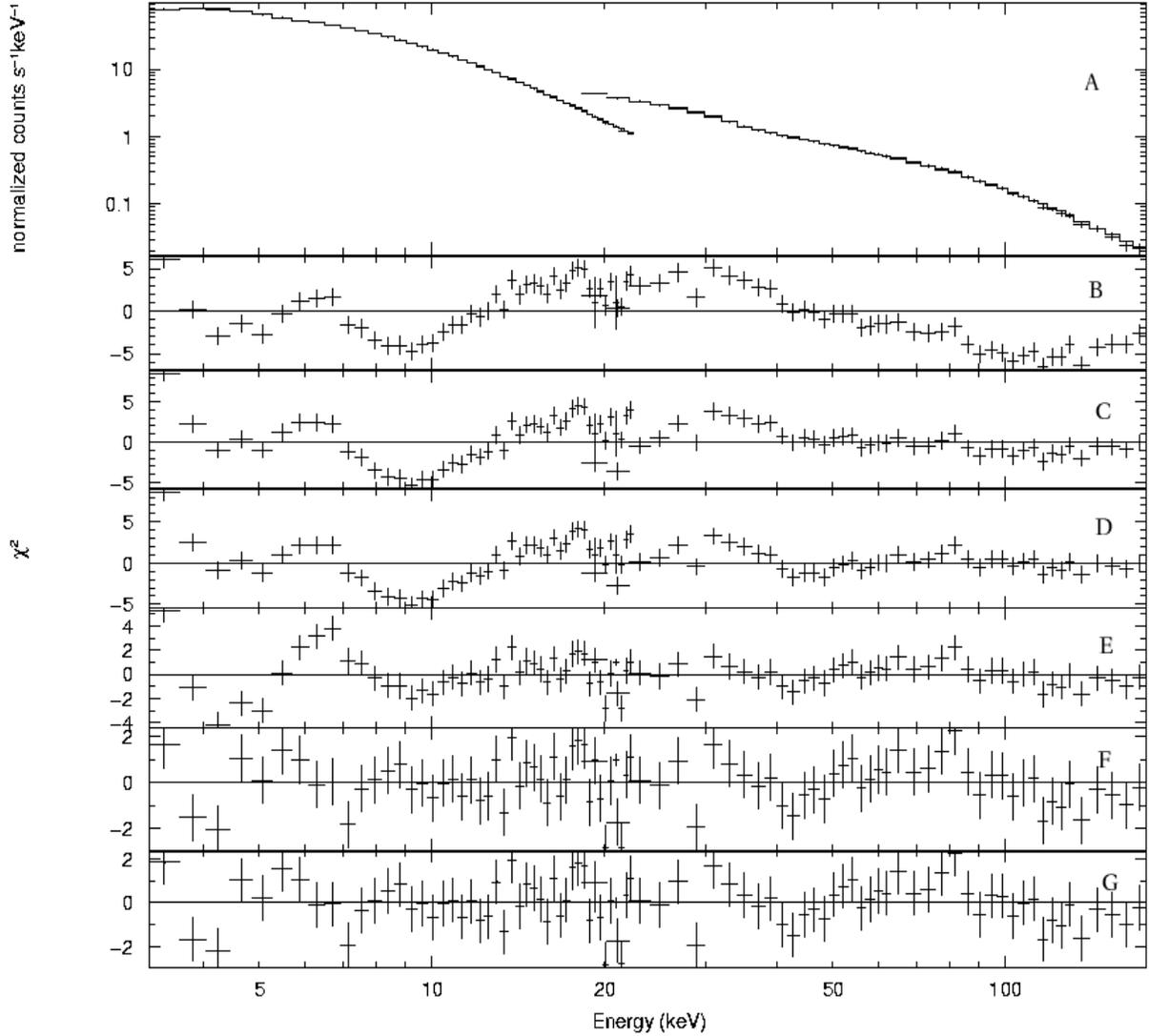}
\caption{Combined PCA/HEXTE spectrum  of {\swifts} for a typical
observation (ID 9143-01-01-00) and the best fit models (panel A).
The residuals are shown subsequently in panel B by taking only the
powerlaw model ($\chi^2$=10.971), panel C the  compTT model
($\chi^2$=6.14), panel D the bknpower model ($\chi^2$=5.10), panel E
the diskbb + bknpower model ($\chi^2$=2.21), panel F the diskbb +
gauss + bknpower model ($\chi^2$=1.16) and panel G the bbody + gauss
+ bknpower model ($\chi^2$=1.19)} \label{phamodel}
\end{figure}

\begin{figure}[ptbptbptb]
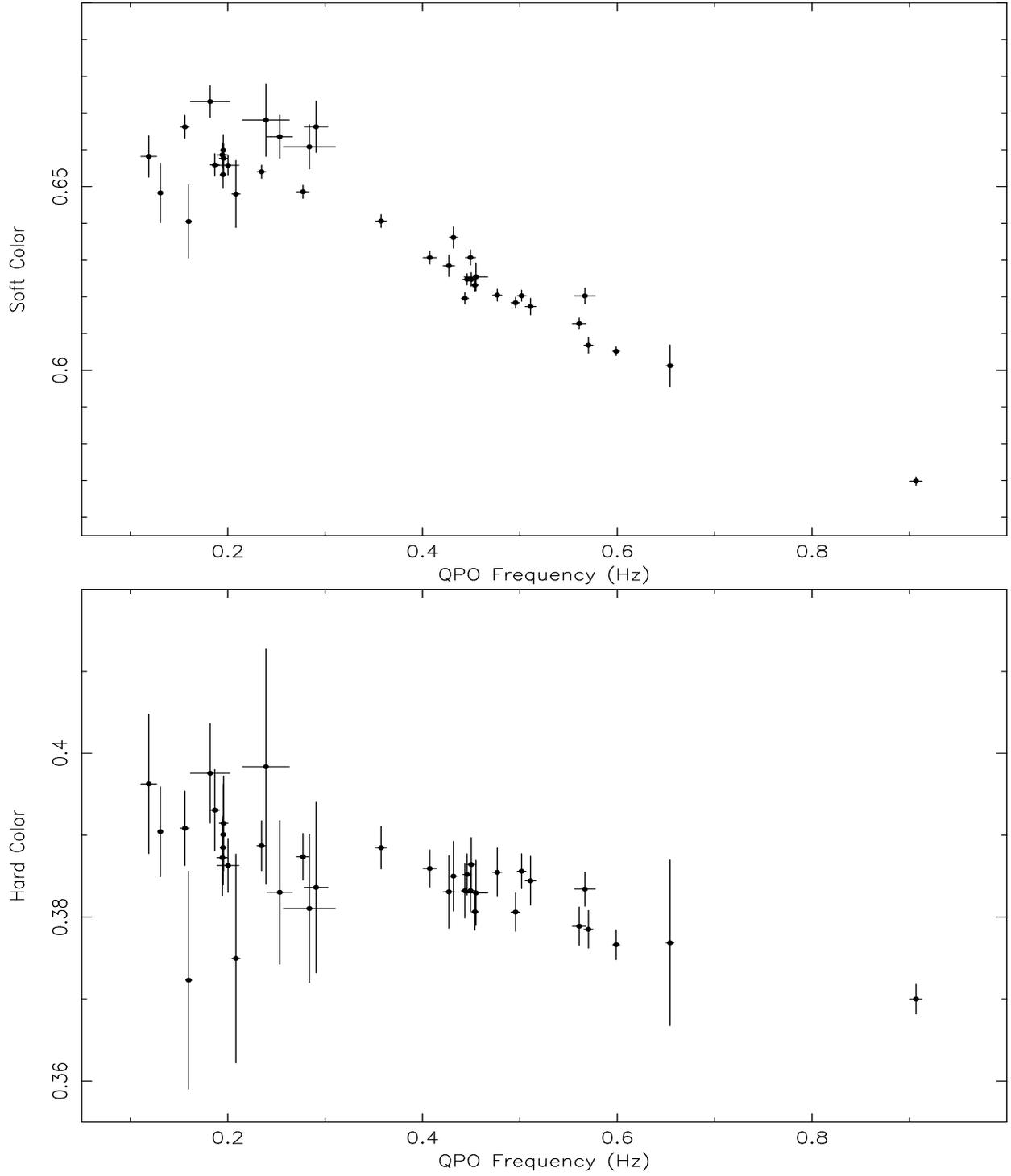

\includegraphics[width=3.8in,height=6.5in,angle=-90]{f7a.eps}
\includegraphics[width=3.8in,height=6.5in,angle=-90]{f7b.eps}
\caption{ Diagrams of the spectral color vs QPO frequency. The upper
panel is for soft color (6.1-9.4 keV/2.1-5.7 keV) and the lower
panel is for hard color (13.5-16.9 keV/9.8-13.1 keV). }
\label{qpo_color}
\end{figure}

We would like to note that in Fig.\ref{qpoflux} there are several
data points staying away from the overall trend of a linear
relationship between the flux and QPO. This phenomenon in
  could not be well explained  by the models of
\cite{markwardt99}, \cite{reig00}, and \cite{titarchuk00}. As shown
in Fig.\ref{lc_color_qpo_time}, an additional small flare occurred
in the third time region is correlated to the three QPO signals
standing out of the lowest flux  in Fig.\ref{qpoflux}. During this
small flare, the spectrum becomes soft (panels D and E in
Fig.\ref{lc_color_qpo_time}). This might suggest  that the QPO
origins from  the perturbation at somewhere of accretion disk, since
otherwise the high QPO frequency would require larger luminosity
than observed according to the main trend of a linear relationship.
The QPO frequency (Fig.\ref{qpoflux}) detected at the peak flux of
the outburst  drops  from 0.906 Hz to 0.654 Hz within one day, while
the  flux decreases by only 9 percent and is largely deviated from
the overall linear trend. This is in general  not consistent with
the arguments of Meyer-Hofmeister (2004) for explaining the
relationship between the QPO frequency and  flux. In the paper of
Meyer-Hofmeister (2004), the transition in spectral state  is the
result of the balance between the evaporation of the inner edge of
  disk and the inward accretion flow. The closer  to the black
hole for the inner disk, the larger the flux and the higher value
the  QPO frequency. In the  region very close to the black hole, the
coronal evaporation becomes dominant, and instead of being
completely truncated the inner edge of the disk might constitute a
soft extension toward the corona. As a result, the source might
remain hard state due to the evaporation of the accretion flow, and
presents the larger frequency of QPO in the power spectrum. The
deviation of this point from the overall trend might provide the
evidence from the observational point of view that the inner edge of
the disk might have a soft extension and evaporate rapidly to
prevent from establishing effectively the high/soft state.

An alternative explanation might be the radial inflow
effect\citep{vander01}. By connecting the deviated QPO points in
Fig.\ref{qpoflux} one gets a line which is roughly parallel to the
track of the main linear trend. Similar parallel tracks are observed
in some kHz-QPO sources in plots of QPO frequency vs. X-ray
luminosity. According to \cite{vander01}, in the so-called radial
inflow scenario, the QPO frequency is determined by some balance
between the accretion rate through disk and radiative stresses, and
it can be expressed as: $\nu(t)\propto\left({\dot M_d(t)\over
L_x(t)} \right)^\beta,(\beta>0)$, where $\dot M_d(t)$ is the
instantaneous accretion rate and $L_x(t)$ is the luminosity.
$L_x(t)$ has both an immediate response to $\dot M_d(t)$ and the
filtered one $\av{\dot M_d}(t)$ averaged over a special time period:
$L_x(t)\propto\dot M_d(t)+\alpha\av{\dot M_d}(t),(\alpha>0)$. The
adjustments on the values of $\dot M_d(t)$ and $\av{\dot M_d}(t)$
could lead to the parallel tracks  in the plot of QPO vs. flux, as
might be the case as well for {\swifts}.

\begin{figure}[ptbptbptb]
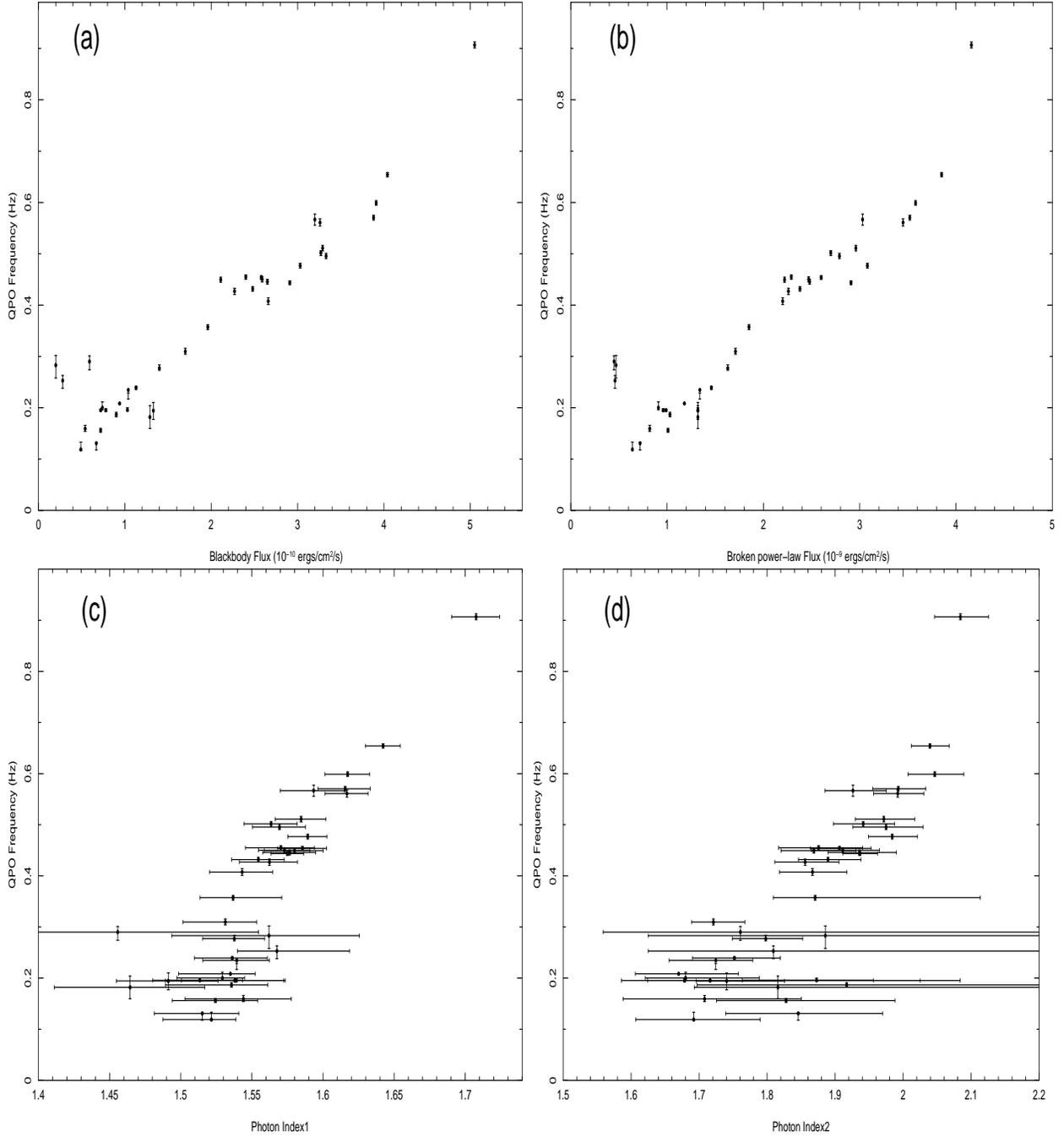

\includegraphics[width=3.5 in,height=3.2in,angle=-90]{f8a.eps}
\includegraphics[width=3.5 in,height=3.2in,angle=-90]{f8b.eps}
\includegraphics[width=3.5 in,height=3.2in,angle=-90]{f8c.eps}
\includegraphics[width=3.5 in,height=3.2in,angle=-90]{f8d.eps}
\caption{ The QPO dependence on the fluxes (a: blackbody; b: broken
power-law) and the spectral indices(c,d), as estimated from the
procedure of spectral fitting.} \label{qpo1_vs_parameter}
\end{figure}

\begin{figure}[ptbptbptb]
\includegraphics[width=5in,height=6in,angle=-90]{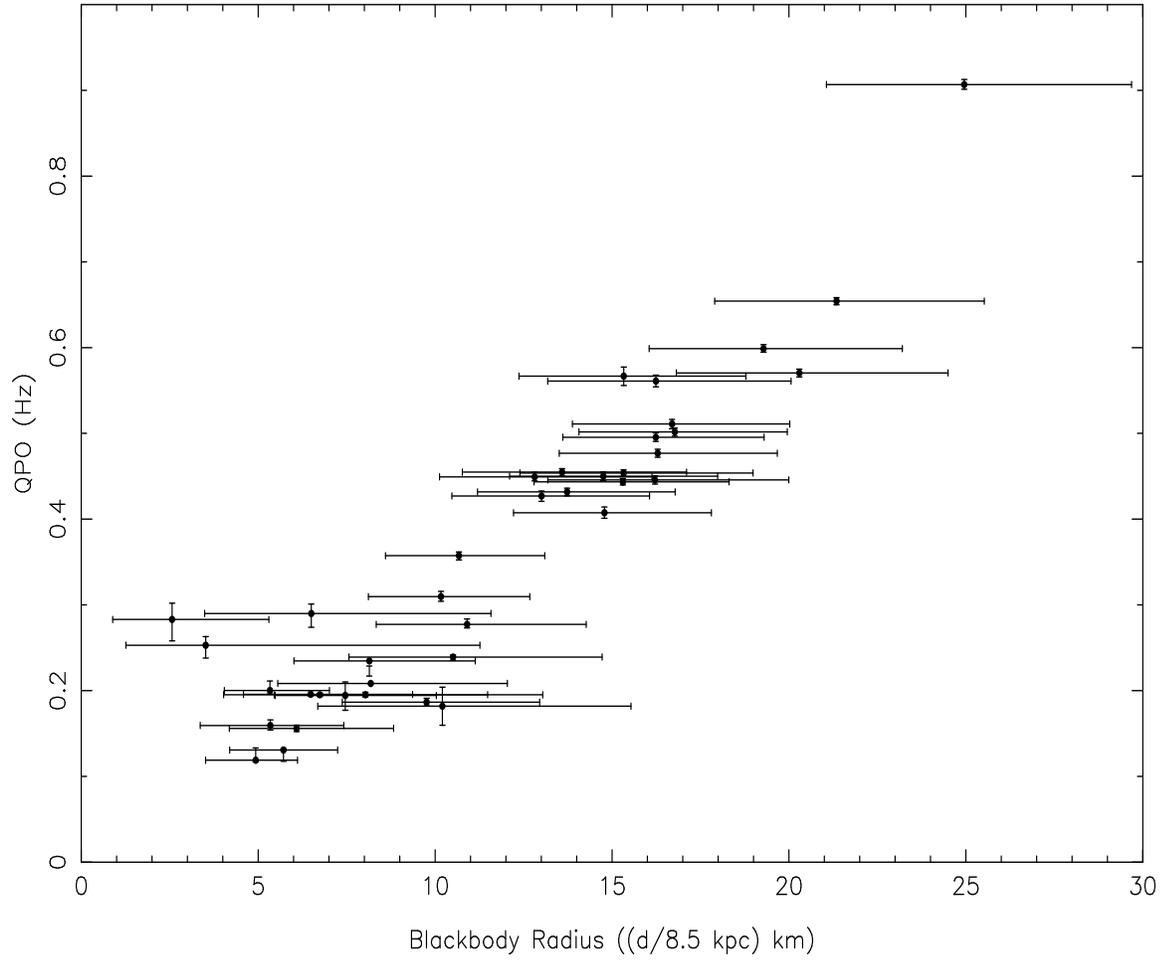}
\caption{ The QPO dependence on the  radius of blackbody region
estimated from the spectral fitting. } \label{qporadius}
\end{figure}

In panel a \& b of Fig.\ref{qpo1_vs_parameter}, we plot the
relations between the QPO frequency and the flux of blackbody and
broken power-law respectively, and find that both of them have a
positive linear relation, implying  the LFQPO  correlation  with the
spectral parameters for both the thermal and nonthermal components.
Similarly,  \cite{markwardt99} and \cite{trudolyubov99} have shown
that the frequency of QPO between 0.5 and 10 Hz of GRS 1915 + 105 is
positively correlated with both the thermal and nonthermal flux.
However, in imagination, the oscillation could  originate in the
accretion disk if the power-law spectrum comes from the inverse
Compton scattering of disk photons,  and so the thermal and
nonthermal components would have some close correlation. Thus this
QPO appears to link both the accretion disk and density of Compton
scattering electrons.  Using the normalization of blackbody model,
we calculate the equivalent radius of blackbody area, and plot its
relation with the QPO frequency in Fig.\ref{qporadius}, showing  a
linear relation. When the flux drops, the equivalent radius gets
smaller, i.e. the radiation area of blackbody in the disk area
decreases, and the QPO frequency decreases because the radius of
inner disk increases.

In the work by \cite{titarchuk04}, they declaim that the observed
low QPO frequency---spectral-index correlation is a natural
consequence of an adjustment of the Keplerian disk flow to the
innermost sub-Keplerian boundary conditions near the central object.
This ultimately leads to the formation of the sub-Keplerian
transition layer between the adjustment radius and the innermost
boundary (the horizon for BH). In panel c and d of
Fig.\ref{qpo1_vs_parameter}, we present the correlation of photon
index $\Gamma$ versus QPO frequency. For QPO frequency values above
0.4 Hz, the QPO frequency and power-law index appear to be
correlated each other positively. However, for the frequency $\leq$
0.4 Hz, the photon power-law index is determined less accurate, and
remains almost constant. The similar properties of the index--QPO
correlation in GRS 1915+105 and XTE J1550-546 have been found
\citep{vignarca03,titarchuk04,shaposhnikov06}. Then in GRS 1915+105,
\cite{titarchuk04} stress that the hard state is related to an
extended Compton cloud (cavity) characterized by a photon index of
around 1.7 and low QPO frequencies below 1 Hz.
 This is the regime where thermal Comptonization dominates the
upscattering of soft disk photons and the spectral shape (index) is
almost independent of mass accretion rate. Thus, panel c and d of
Fig.\ref{qpo1_vs_parameter} suggest that the   source was in the
low/hard state. In addition, \cite{shaposhnikov06} argues that the
index saturation level is determined by the temperature of the
converging flow where the soft (disk) photons are upscattered by
electrons to the energy of falling electrons. In principle, one can
evaluate the mass of the central black hole using the index-QPO
relation because QPO frequencies are inversely proportional to the
BH mass \citep{titarchuk04,shaposhnikov06}.

\section{ CONCLUSION}

Almost all the lights under the current analysis point to the fact
that {\swifts} kept staying in the low/hard state during the
outburst, along with the evolutions of  the spectral and the timing
features similar to those of  other BHCs like XTE J1118+480 and GRO
J0422+32. By far all these sources belong to this small subset of
BHC were observed to have almost only the low/hard state while in
outburst. The corresponding scenario could be that, due to the low
peak luminosity, the coronal evaporation truncates the thin disk
even in outburst and the spectra have to stay hard \citep{meyer04}.
During the outburst, {\swifts} was detected to have its QPO
frequency varying  linearly with the flux for most of the time. An
interesting deviation from this trend as observed at the peak flux
might provide the first evidence of 'seeing' a rapid evaporation for
the inner edge of the disk, which prevents from effectively
establishing the high/soft state during the early stage of the
flare.  The relationship between the QPO frequency and spectral
color is anti-correlation. Although on a theoretical ground the
mechanism of the LFQPO of {\swifts} is still open, this seems not
likely to be satisfied  by the global disk oscillation model
\citep{titarchuk00}.

\acknowledgments This work has been partially supported by the
National Nature Science Foundation of China , Hebei Province Nature
Science Fundamental Project No.A2006000128 and made use of data
obtained through the HEASARC online service, provided by the
NASA/GSFC.

\newpage

\newpage

\newpage

\newpage

\newpage

\newpage

\newpage

\newpage

\centering
\begin{table*}
\begin{center}
\caption{The QPO frequency  as obtained for each dataset. } \small
\begin{tabular}{lcccccc}

\hline \hline
Obs. ID & Start Data & exposure (s) & QPO (Hz) \\
%&&(S)&(Hz)\\
\hline
 91423-01-01-04 & 2005-07-06T05:13:20 & 3232 & $0.906^{+0.006}_{-0.005}$    \\
 91423-01-01-00 & 2005-07-07T04:49:03 & 6464 & $0.654^{+0.004}_{-0.005}$    \\
 91423-01-02-00 & 2005-07-12T01:13:20 & 3232 & $0.598^{+0.005}_{-0.004}$    \\
 91423-01-02-05 & 2005-07-12T12:14:24 & 3232 & $0.570^{+0.003}_{-0.005}$    \\
 91423-01-02-06 & 2005-07-13T21:17:20 & 3232 & $0.560^{+0.007}_{-0.006}$    \\
 91423-01-03-00 & 2005-07-16T01:11:28 & 5264 & $0.476^{+0.005}_{-0.004}$    \\
 91423-01-03-02 & 2005-07-17T13:22:47 & 3216 & $0.510^{+0.006}_{-0.005}$    \\
 91423-01-03-03 & 2005-07-18T17:41:20 & 3204 & $0.566^{+0.011}_{-0.009}$    \\
 91423-01-03-04 & 2005-07-19T12:34:24 & 3216 & $0.495^{+0.005}_{-0.005}$    \\
 91423-01-03-05 & 2005-07-20T12:09:20 & 3136 & $0.501^{+0.005}_{-0.005}$    \\
 91423-01-03-06 & 2005-07-21T21:12:12 & 3200 & $0.453^{+0.004}_{-0.004}$    \\
 91423-01-03-07 & 2005-07-16T23:13:32 & 8352 & $0.443^{+0.004}_{-0.004}$    \\
 91423-01-04-00 & 2005-07-22T20:47:28 & 3200 & $0.445^{+0.004}_{-0.005}$    \\
 91423-01-04-01 & 2005-07-23T15:39:28 & 3184 & $0.450^{+0.004}_{-0.005}$    \\
 91423-01-04-02 & 2005-07-24T12:06:24 & 3200 & $0.431^{+0.005}_{-0.004}$    \\
 91423-01-04-03 & 2005-07-25T18:06:58 & 2800 & $0.407^{+0.007}_{-0.006}$    \\
 91423-01-04-04 & 2005-07-26T17:35:28 & 3200 & $0.427^{+0.005}_{-0.007}$    \\
 91423-01-04-05 & 2005-07-27T04:35:47 & 1760 & $0.454^{+0.004}_{-0.004}$    \\
 91423-01-04-06 & 2005-07-28T10:28:16 & 2928 & $0.449^{+0.005}_{-0.005}$    \\
 91423-01-05-01 & 2005-08-01T15:09:18 & 3200 & $0.357^{+0.005}_{-0.004}$    \\
 91423-01-05-02 & 2005-08-03T15:55:51 & 3200 & $0.309^{+0.004}_{-0.005}$    \\
 91423-01-06-00 & 2005-08-05T15:06:40 & 3216 & $0.277^{+0.005}_{-0.004}$    \\
 91423-01-06-01 & 2005-08-07T11:15:27 & 2896 & $0.239^{+0.002}_{-0.004}$    \\
 91423-01-06-06 & 2005-08-09T18:44:32 & 1344 & $0.194^{+0.016}_{-0.017}$    \\
 91423-01-06-07 & 2005-08-09T22:00:32 & 944  & $0.181^{+0.022}_{-0.022}$    \\
 91423-01-06-03 & 2005-08-11T04:51:28 & 3104 & $0.234^{+0.007}_{-0.017}$    \\
 91423-01-07-00 & 2005-08-13T17:01:20 & 1680 & $0.195^{+0.005}_{-0.002}$    \\
 91423-01-07-01 & 2005-08-15T06:45:52 & 1680 & $0.208^{+0.004}_{-0.005}$    \\
 91423-01-08-00 & 2005-08-19T09:25:36 & 3216 & $0.155^{+0.003}_{-0.994}$    \\
 91423-01-08-01 & 2005-08-21T08:35:44 & 3200 & $0.186^{+0.005}_{-0.004}$    \\
 91423-01-08-02 & 2005-08-23T11:03:12 & 1120 & $0.195^{+0.001}_{-0.002}$    \\
 91423-01-08-03 & 2005-08-25T05:53:52 & 1472 & $0.195^{+0.003}_{-0.003}$    \\
 91423-01-09-00 & 2005-08-30T17:39:44 & 2800 & $0.200^{+0.011}_{-0.004}$    \\
 91423-01-10-00 & 2005-09-06T16:17:04 & 1968 & $0.159^{+0.003}_{-0.004}$    \\
 91423-01-11-00 & 2005-09-10T17:46:40 & 3200 & $0.130^{+0.002}_{-0.013}$    \\
 91423-01-11-01 & 2005-09-14T08:17:20 & 3200 & $0.118^{+0.003}_{-0.002}$    \\
 91423-01-16-00 & 2005-10-15T08:15:28 & 944  & $0.290^{+0.011}_{-0.016}$    \\
 91423-01-16-01 & 2005-10-19T18:06:08 & 1248 & $0.283^{+0.018}_{-0.025}$    \\
 91423-01-17-00 & 2005-10-22T13:29:52 & 300  & $0.253^{+0.010}_{-0.015}$    \\
\hline \label{qpotable}

\end{tabular}
\end{center}
\end{table*}

\centering
\begin{table*}
\begin{center}
\tiny \caption{Comparison of the fit by applying different models on
the combined PCA/HEXTE data. }

\begin{tabular}{ccccccccccc}

\hline \hline
model & kT & k$T_c$ & $T0_c$ & Tin &$\Gamma$ & $\Gamma_1$ & $\Gamma_2$ & breakE  & $\chi^2$\\
&(kev)&(keV)&(keV)&(keV)&&&&(keV)&\\
\hline
pow               & --             & --                   & --             &  --            & $1.76\pm 0.02$ & --             & --             & --                     & 10.97 \\
compTT            & --             & $178^{+\ 9}_{-12}$ & $0.45\pm 0.03$   &             -- & --             & --             & --             & --                     & 6.14 \\
bknpo             & --             & --                   & --             & --             & --             & $1.74\pm 0.02$ & $2.03\pm 0.05$ & $46.8^{+ 3.5}_{- 3.2}$ & 5.10 \\
diskbb+bknpo      & --             & --                   & --             & $1.13\pm 0.06$ & --             & $1.63\pm 0.02$ & $2.03\pm 0.03$ & $35.8^{+ 1.9}_{- 1.1}$ & 2.21 \\
diskbb+gau+bknpo  & --             & --                   & --             & $0.86\pm 0.04$ & --             & $1.64\pm 0.01$ & $2.04\pm 0.03$ & $36.0^{+ 2.3}_{- 2.1}$ & 1.16 \\
bb+gau+bknpo      & $0.66\pm 0.04$ & --                   & --             &  --            & --             & $1.64\pm 0.01$ & $2.04\pm 0.03$ & $36.4^{+ 1.7}_{- 1.7}$ & 1.19 \\

\hline \label{tab:model}

\end{tabular}
\end{center}
\end{table*}

\centering
\begin{table*}
\begin{center}
\tiny \caption{Parameters (the 3-6th columns) from fitting the data
(the first column shows the data ID, and the second column shows the
day when the observation was carried out by the model of bbody +
gauss + bknpower. The 7th and 8th columns present the estimated
luminosity of the soft component (blackbody) and hard component
(broken power-law). The reduced $\chi^2$ in the last column describe
the quality of each fit.}

\begin{tabular}{ccccccccccc}

\hline \hline
Obs. ID & MJD & kT & $radius_{bb}$ & $\Gamma_1$ & $\Gamma_2$ & breakE & EW & $\rm flux_{bb}$ & $\rm flux_{bkn}$ & $\chi^2$\\
&&(keV)&(d/8.5 kpc) km&&&keV&eV&$10^{-10}$ergs/$cm^2$/s&$10^{-9}$ergs/$cm^2$/s&\\
\hline
91423-01-01-04  & 53557 & $0.65^{+0.04}_{-0.04}$ & $24.9^{+4.7}_{-3.8}$ & $1.71^{+0.02}_{-0.02}$ & $2.08^{+0.04}_{-0.04}$ & $35.3^{+ 2.5}_{- 2.3}$ & 177 & 5.05 & 4.16 & 1.19 \\
91423-01-01-00  & 53558 & $0.66^{+0.04}_{-0.04}$ & $21.3^{+4.1}_{-3.4}$ & $1.64^{+0.01}_{-0.01}$ & $2.04^{+0.03}_{-0.03}$ & $36.3^{+ 1.7}_{- 1.6}$ & 152 & 4.04 & 3.85 & 1.20 \\
91423-01-02-00  & 53563 & $0.69^{+0.05}_{-0.05}$ & $19.2^{+3.9}_{-3.2}$ & $1.62^{+0.02}_{-0.02}$ & $2.05^{+0.04}_{-0.04}$ & $36.8^{+ 2.4}_{- 2.2}$ & 144 & 3.91 & 3.58 & 1.55 \\
91423-01-02-05  & 53563 & $0.67^{+0.05}_{-0.05}$ & $20.2^{+4.2}_{-3.4}$ & $1.62^{+0.02}_{-0.02}$ & $1.99^{+0.04}_{-0.04}$ & $33.0^{+ 2.3}_{- 2.3}$ & 162 & 3.88 & 3.52 & 1.24 \\
91423-01-02-06  & 53564 & $0.71^{+0.06}_{-0.05}$ & $16.2^{+3.8}_{-3.0}$ & $1.62^{+0.01}_{-0.02}$ & $1.99^{+0.04}_{-0.04}$ & $35.5^{+ 2.5}_{- 2.3}$ &  97 & 3.26 & 3.45 & 1.51 \\
91423-01-03-00  & 53567 & $0.70^{+0.05}_{-0.05}$ & $16.2^{+3.3}_{-2.7}$ & $1.59^{+0.01}_{-0.01}$ & $1.98^{+0.04}_{-0.03}$ & $38.4^{+ 2.4}_{- 2.3}$ & 131 & 3.03 & 3.08 & 1.50 \\
91423-01-03-07  & 53567 & $0.71^{+0.05}_{-0.05}$ & $15.3^{+3.0}_{-2.5}$ & $1.58^{+0.01}_{-0.01}$ & $1.94^{+0.03}_{-0.02}$ & $37.0^{+ 1.7}_{- 1.7}$ & 122 & 2.91 & 2.91 & 1.56 \\
91423-01-03-02  & 53568 & $0.70^{+0.05}_{-0.05}$ & $16.7^{+3.3}_{-2.8}$ & $1.58^{+0.02}_{-0.02}$ & $1.97^{+0.05}_{-0.04}$ & $35.5^{+ 2.6}_{- 2.5}$ & 157 & 3.29 & 2.96 & 1.09 \\
91423-01-03-03  & 53569 & $0.72^{+0.06}_{-0.05}$ & $15.3^{+3.4}_{-2.9}$ & $1.59^{+0.02}_{-0.02}$ & $1.93^{+0.05}_{-0.04}$ & $31.4^{+ 3.8}_{- 3.2}$ & 155 & 3.20 & 2.03 & 1.22 \\
91423-01-03-04  & 53570 & $0.71^{+0.05}_{-0.05}$ & $16.2^{+3.0}_{-2.6}$ & $1.57^{+0.02}_{-0.02}$ & $1.97^{+0.05}_{-0.05}$ & $38.0^{+ 3.3}_{- 2.9}$ & 180 & 3.33 & 2.79 & 0.95 \\
91423-01-03-05  & 53571 & $0.70^{+0.05}_{-0.04}$ & $16.7^{+3.1}_{-2.7}$ & $1.56^{+0.02}_{-0.02}$ & $1.94^{+0.05}_{-0.04}$ & $34.7^{+ 2.6}_{- 2.6}$ & 183 & 3.27 & 2.70 & 1.12 \\
91423-01-03-06  & 53572 & $0.69^{+0.06}_{-0.05}$ & $15.3^{+3.6}_{-2.9}$ & $1.59^{+0.02}_{-0.02}$ & $1.91^{+0.05}_{-0.04}$ & $36.8^{+ 3.5}_{- 3.2}$ & 119 & 2.58 & 2.60 & 1.24 \\
91423-01-04-00  & 53573 & $0.68^{+0.05}_{-0.05}$ & $16.2^{+3.7}_{-3.0}$ & $1.58^{+0.02}_{-0.02}$ & $1.94^{+0.05}_{-0.05}$ & $37.1^{+ 3.9}_{- 3.2}$ & 157 & 2.65 & 2.48 & 1.45 \\
91423-01-04-01  & 53574 & $0.70^{+0.05}_{-0.05}$ & $14.7^{+3.2}_{-2.6}$ & $1.57^{+0.02}_{-0.02}$ & $1.91^{+0.05}_{-0.05}$ & $36.8^{+ 4.2}_{- 3.4}$ & 137 & 2.59 & 2.47 & 0.97 \\
91423-01-04-02  & 53575 & $0.72^{+0.06}_{-0.05}$ & $13.7^{+3.0}_{-2.5}$ & $1.55^{+0.02}_{-0.02}$ & $1.89^{+0.05}_{-0.04}$ & $36.1^{+ 3.5}_{- 3.2}$ & 124 & 2.48 & 2.38 & 1.30 \\
91423-01-04-03  & 53576 & $0.71^{+0.05}_{-0.05}$ & $14.7^{+3.0}_{-2.5}$ & $1.54^{+0.02}_{-0.02}$ & $1.87^{+0.05}_{-0.05}$ & $31.4^{+ 3.8}_{- 4.3}$ & 162 & 2.66 & 2.20 & 0.93 \\
91423-01-04-04  & 53577 & $0.72^{+0.06}_{-0.06}$ & $13.0^{+3.0}_{-2.5}$ & $1.56^{+0.02}_{-0.02}$ & $1.86^{+0.05}_{-0.04}$ & $33.5^{+ 4.2}_{- 3.9}$ & 132 & 2.27 & 2.26 & 0.94 \\
91423-01-04-05  & 53578 & $0.72^{+0.07}_{-0.06}$ & $13.5^{+3.5}_{-2.8}$ & $1.57^{+0.02}_{-0.02}$ & $1.88^{+0.06}_{-0.06}$ & $34.3^{+ 4.7}_{- 4.8}$ & 116 & 2.40 & 2.29 & 0.98 \\
91423-01-04-06  & 53579 & $0.71^{+0.07}_{-0.06}$ & $12.8^{+3.3}_{-2.6}$ & $1.58^{+0.02}_{-0.02}$ & $1.87^{+0.06}_{-0.05}$ & $36.6^{+ 5.9}_{- 4.1}$ & 137 & 2.11 & 2.22 & 1.34 \\
91423-01-05-01  & 53583 & $0.76^{+0.07}_{-0.06}$ & $10.6^{+2.4}_{-2.0}$ & $1.54^{+0.03}_{-0.02}$ & $1.87^{+0.24}_{-0.06}$ & $39.4^{+19.8}_{- 5.4}$ & 145 & 1.96 & 1.85 & 1.26 \\
91423-01-05-02  & 53585 & $0.75^{+0.07}_{-0.06}$ & $10.1^{+2.5}_{-2.0}$ & $1.53^{+0.02}_{-0.03}$ & $1.72^{+0.05}_{-0.03}$ & $23.1^{+ 7.5}_{- 5.9}$ & 127 & 1.70 & 1.71 & 1.12 \\
91423-01-06-00  & 53587 & $0.70^{+0.08}_{-0.07}$ & $10.9^{+3.3}_{-2.5}$ & $1.54^{+0.02}_{-0.02}$ & $1.80^{+0.05}_{-0.05}$ & $34.7^{+ 4.8}_{- 4.7}$ & 144 & 1.40 & 1.63 & 0.85 \\
91423-01-06-01  & 53589 & $0.68^{+0.09}_{-0.08}$ & $10.5^{+4.2}_{-2.9}$ & $1.54^{+0.02}_{-0.03}$ & $1.75^{+0.07}_{-0.06}$ & $34.8^{+ 8.7}_{- 8.5}$ & 134 & 1.13 & 1.46 & 1.03 \\
91423-01-06-05  & 53590 & $0.87^{+0.13}_{-0.12}$ & $ 6.1^{+2.3}_{-2.1}$ & $1.52^{+0.04}_{-0.04}$ & $1.74^{+0.17}_{-0.11}$ & $39.1^{+23.1}_{-18.9}$ &  79 & 1.21 & 1.42 & 0.97 \\
91423-01-06-04  & 53591 & $0.72^{+0.15}_{-0.15}$ & $ 7.6^{+4.0}_{-1.8}$ & $1.56^{+0.03}_{-0.03}$ & $2.50^{+0.52}_{-0.40}$ & $79.5^{+ 9.2}_{-12.7}$ &  78 & 0.77 & 1.44 & 1.15 \\
91423-01-06-06  & 53591 & $0.81^{+0.11}_{-0.05}$ & $ 7.4^{+2.5}_{-1.9}$ & $1.49^{+0.03}_{-0.04}$ & $1.74^{+0.08}_{-0.04}$ & $31.6^{+ 6.7}_{- 7.4}$ & 131 & 1.33 & 1.32 & 1.18 \\
91423-01-06-07  & 53591 & $0.71^{+0.12}_{-0.12}$ & $10.2^{+5.3}_{-3.5}$ & $1.46^{+0.05}_{-0.05}$ & $1.82^{+8.18}_{-0.12}$ & $35.0^{+27.2}_{- 8.1}$ & 180 & 1.29 & 1.32 & 1.08 \\
91423-01-06-03  & 53593 & $0.74^{+0.10}_{-0.09}$ & $ 8.1^{+2.9}_{-2.1}$ & $1.54^{+0.02}_{-0.02}$ & $1.72^{+0.05}_{-0.07}$ & $34.9^{+ 8.0}_{-23.6}$ &  98 & 1.04 & 1.34 & 1.05 \\
91423-01-08-03  & 53593 & $0.71^{+0.13}_{-0.13}$ & $ 8.0^{+5.0}_{-2.5}$ & $1.54^{+0.03}_{-0.04}$ & $1.68^{+0.35}_{-0.09}$ & $32.5^{+46.3}_{-21.4}$ &  89 & 0.78 & 0.96 & 0.87 \\
91423-01-07-00  & 53595 & $0.82^{+0.14}_{-0.12}$ & $ 6.4^{+2.8}_{-1.9}$ & $1.51^{+0.03}_{-0.03}$ & $1.87^{+0.08}_{-0.11}$ & $45.8^{+11.0}_{- 7.7}$ &  68 & 1.03 & 1.32 & 0.75 \\
91423-01-07-01  & 53597 & $0.73^{+0.12}_{-0.11}$ & $ 8.1^{+3.8}_{-2.6}$ & $1.53^{+0.02}_{-0.04}$ & $1.67^{+0.09}_{-0.06}$ & $29.9^{+14.9}_{-14.3}$ &  85 & 0.94 & 1.18 & 0.89 \\
91423-01-08-00  & 53601 & $0.78^{+0.13}_{-0.12}$ & $ 6.0^{+2.7}_{-1.9}$ & $1.52^{+0.03}_{-0.03}$ & $1.83^{+0.16}_{-0.10}$ & $41.4^{+13.8}_{-10.4}$ &  97 & 0.72 & 1.01 & 1.12 \\
91423-01-08-01  & 53603 & $0.67^{+0.08}_{-0.08}$ & $ 9.7^{+3.2}_{-2.3}$ & $1.54^{+0.03}_{-0.05}$ & $1.92^{+0.93}_{-0.22}$ & $54.2^{+34.7}_{-24.7}$ & 151 & 0.90 & 1.03 & 1.17 \\
91423-01-08-02  & 53605 & $0.75^{+0.17}_{-0.15}$ & $ 6.7^{+4.7}_{-2.7}$ & $1.54^{+0.03}_{-0.05}$ & $1.72^{+0.37}_{-0.09}$ & $32.2^{+50.9}_{-16.4}$ &  96 & 0.72 & 0.99 & 0.70 \\
91423-01-09-00  & 53612 & $0.83^{+0.11}_{-0.10}$ & $ 5.3^{+1.6}_{-1.2}$ & $1.53^{+0.02}_{-0.03}$ & $1.68^{+0.11}_{-0.06}$ & $26.2^{+15.4}_{- 9.1}$ &  80 & 0.74 & 0.91 & 0.83 \\
91423-01-10-00  & 53623 & $0.77^{+0.18}_{-0.16}$ & $ 5.3^{+2.0}_{-1.9}$ & $1.54^{+0.03}_{-0.04}$ & $1.71^{+0.14}_{-0.12}$ & $36.4^{+22.5}_{-21.7}$ &  86 & 0.54 & 0.82 & 1.13 \\
91423-01-11-00  & 53627 & $0.79^{+0.11}_{-0.05}$ & $ 5.7^{+1.5}_{-1.5}$ & $1.52^{+0.03}_{-0.03}$ & $1.85^{+0.12}_{-0.11}$ & $39.1^{+ 7.9}_{- 7.4}$ & 121 & 0.67 & 0.72 & 0.88 \\
91423-01-11-01  & 53630 & $0.78^{+0.06}_{-0.12}$ & $ 4.9^{+1.1}_{-1.4}$ & $1.52^{+0.02}_{-0.03}$ & $1.69^{+0.10}_{-0.09}$ & $30.5^{+13.5}_{-11.5}$ &  78 & 0.49 & 0.64 & 1.01 \\
91423-01-12-00  & 53634 & $0.88^{+0.13}_{-0.15}$ & $ 3.3^{+1.4}_{-0.9}$ & $1.54^{+0.03}_{-0.04}$ & $1.75^{+0.13}_{-0.08}$ & $29.1^{+13.6}_{- 9.7}$ &  42 & 0.39 & 0.60 & 0.93 \\
91423-01-12-01  & 53638 & $0.74^{+0.13}_{-0.13}$ & $ 5.6^{+2.8}_{-1.8}$ & $1.51^{+0.03}_{-0.07}$ & $1.68^{+0.07}_{-0.06}$ & $17.9^{+19.2}_{- 4.2}$ & 111 & 0.49 & 0.55 & 0.98 \\
group1$^*$      & ----- & $0.84^{+0.09}_{-0.04}$ & $ 3.9^{+0.9}_{-0.6}$ & $1.48^{+0.03}_{-0.03}$ & $1.74^{+0.11}_{-0.12}$ & $30.5^{+ 9.2}_{- 8.0}$ &  85 & 0.48 & 0.48 & 1.37 \\
91423-01-16-00  & 53658 & $0.73^{+0.22}_{-0.17}$ & $ 6.5^{+4.3}_{-2.5}$ & $1.45^{+0.10}_{-0.20}$ & $1.76^{+4.22}_{-0.11}$ & $39.4^{+60.3}_{-19.3}$ & 213 & 0.59 & 0.45 & 1.03 \\
91423-01-16-01  & 53662 & $0.92^{+0.32}_{-0.35}$ & $ 2.4^{+2.3}_{-1.4}$ & $1.56^{+0.06}_{-0.07}$ & $1.88^{+0.94}_{-0.09}$ & $46.5^{+45.3}_{-35.4}$ &  79 & 0.20 & 0.47 & 0.78 \\
91423-01-17-00  & 53665 & $0.76^{+0.34}_{-0.73}$ & $ 3.2^{+6.5}_{-1.9}$ & $1.57^{+0.05}_{-0.03}$ & $1.81^{+0.66}_{-0.08}$ & $41.7^{+46.2}_{-27.8}$ &  67 & 0.28 & 0.46 & 1.07 \\
group2$^*$      & ----- & $0.85^{+0.10}_{-0.09}$ & $ 2.8^{+0.8}_{-1.2}$ & $1.54^{+0.02}_{-0.03}$ & $1.77^{+0.08}_{-0.06}$ & $33.5^{+ 7.4}_{- 7.4}$ &  85 & 0.29 & 0.38 & 0.82 \\

\hline \label{tab:bknpo}

\end{tabular}
\vspace{0.2cm}
\parbox{6.8in}
{\baselineskip 5pt {\sc Note~} $^*$~group1 contain Obs. ID
(91423-01-13-00, 91423-01-13-01, 91423-01-14-00, 91423-01-14-01,
91423-01-15-00, 91423-01-15-01). group2 contain Obs. ID
(91423-01-17-01,
91423-01-18-00,91423-01-19-00,91423-01-19-01,91423-01-20-00,
91423-01-20-01,91423-01-21-00,91423-01-21-01,91423-01-22-00,
91423-01-22-01,92404-01-02-00).\\}
% \hspace{0.5cm}$^2$~}
\end{center}
\end{table*}

\end{document}